\documentclass[twocolumn, showpacs,preprintnumbers,amsmath,amssymb]{revtex4} %preprint, draft
\usepackage{graphicx}
\usepackage{amsmath}
\usepackage{amsfonts}
\usepackage{subfigure}
\usepackage{epstopdf}
\begin{document}

\title{Diffusion-free ultrafast carrier dynamics in silicon nano-pillars}

\author{D. Chekulaev\footnote{Corresponding author: dimitri@nprl.ph.bham.ac.uk} and A. Kaplan}
\affiliation{Nanoscale Physics Research Laboratory, School of Physics and Astronomy, University of Birmingham, Birmingham B15 2TT, UK}

\date{\today}
\begin{abstract}
We have investigated ultrafast carriers dynamics in crystalline silicon nano-pillars structure using a pump-probe reflectivity method with 800 nm, 150 fs laser pulses and fluence in the range of $17 \div 170$ mJ/cm$^2$. Dimensions of the structure allow us to eliminate contribution from the diffusion process to the relaxation dynamics of the excited carriers. Strong intensity-dependent time-resolved reflectivity change, $\Delta R(\tau_d)$, was monitored in the submelting regime. At strong pumping $\Delta R(\tau_d)$ can reach up to 8\%, a several times higher than for a bulk silicon. From the measurements  we deduced recombination time of up to 10 ns at carriers density of $\sim 1\times 10^{20}$ cm$^{-3}$, while electron-phonon interaction occurs during 350--400 fs  and it is independent of the concentration.

\end{abstract}


\maketitle
Since the femtosecond laser technology became available to a wide range of research areas,  the subpicosecond carriers dynamics in semiconductors has been subjected to intensive theoretical \cite{Sernelius} and experimental \cite{Shank, Sokolowski, Othonos} research in bulk materials. In recent years considerable attention has been devoted to the development of applications based on Si nanostructures (NS). For example,  Si field emitters based on nano-wires  \cite{Frederick}, numerous applications of NS in photonics: electroluminescence diode, microcavity and waveguide;  electronics: single-electron device and nonvolatile memory  \cite{Koshida}. An understanding of carriers dynamics is crucial for the research and development of such devices.

 Transient measurements of reflectivity and transmissivity, provided by the pump-probe technique, allow temporal monitoring of  changes in the optical properties of materials contributed by excited electrons (and holes) dynamics. The main parameters governing carriers dynamics are: carrier-carrier scattering, occurring on a few femtoseconds time scale \cite{Buhleier}; carrier-phonon thermalization lasting a few hundred femtoseconds \cite{Doany}; and electron-hole (e-h) surface 
 \cite{Sabbah2000} and bulk \cite{Esser} recombinations typically characterized by much longer time scales \cite{Woerdman, Tanaka}.  
 
The drawback estimating of these parameters from the optical measurements is a carrier diffusion from the excited area or 'probing spot', especially for high density e-h plasma, $N_{e-h}> 10^{19}$ cm$^{-3}$ \cite{Li, Tanaka}. This complicates interpretation of results owing to a convolution between diffusion and various carrier-carrier and carrier-phonon interactions \cite{Othonos}. With the development of nanotechnology it has become possible to construct a structure where spatial confinement suppresses the diffusion process though optical properties remain similar to those of the bulk. Such a structure has to be smaller than the diffusion length (estimated as  $\sqrt{\tau_d D_N} \approx 1-100 ~\mu \textrm{m}$, with the diffusion coefficient $D_N=10~\textrm{cm}^2/\textrm{s}$ \cite{Tanaka}) but larger than electron wavelength to avoid effects of the quantum confinement. Thus structures with dimensions of about  100 nm are suitable for the purpose.
 
We present ultrafast transient reflectivity measurements of a crystalline Si nano-pillars (NP) structure where individual Si NP are sufficiently separated from the bulk substrate and each other. Thus, pump-probe measurements of an ensemble of NP represent individual properties of a single pillar free of contributions from  its surrounding.  In this Letter, we show: (i) carrier dynamics below and at the melting threshold  (ii) dependence of the decay constants on the carriers concentration (iii) linear response to the excitation and consequently (iv) utilization of the Drude model. We also observed that the absolute change in the reflectivity is up to one order of magnitude larger than similar changes observed in bulk Si,  and the carriers recombination time is concentration-dependent occurring on a timescale of a  few nanoseconds while thermalization time is about a few hundred femtoseconds independent of the carriers concentration.
\begin{figure}[tbh]
\centering\resizebox{\columnwidth}{!}{\rotatebox{0}{\includegraphics{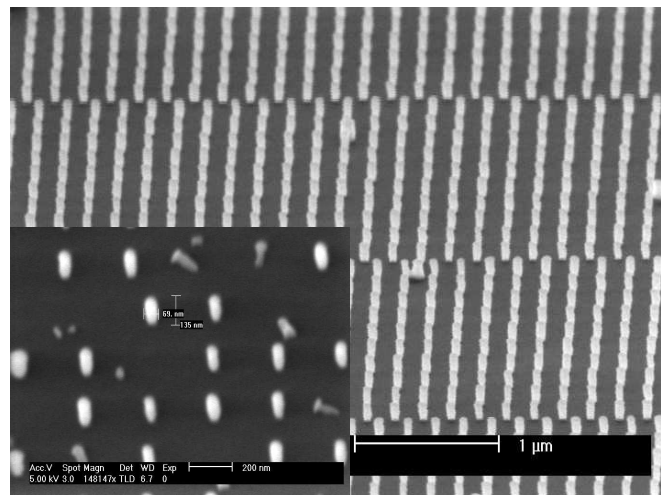}}}
\caption{\label{fig:1} SEM pictures of silicon nanopillars made by imprint method on Silicon-on-Insulator (SOI) wafer. $50^{\circ}$ tilted view with resolution of $1~\mu$m. Inset: $45^{\circ}$ tilted view with resolution of $200$~nm, pillar size estimation: $70\times200$~nm }
\end{figure}

The femtosecond pump-probe reflectivity measurements were performed by use of Ti:sapphire laser oscillator amplified up to 1 mJ per pulse at 800 nm and rep rate of 1 kHz. Second harmonic generation of BBO crystal was used to determine the pulse duration of $\tau_p=150$ fs of nearly Gaussian shape. The pump and probe beams were \textit{p}- and \textit{s}-polarized respectively. Both beams had been nearly normal incident on to the sample (with a small skew angle with respect to each other) focused to a round spot with a diameter of 70 $\mu$m. Careful overlap of the pump and probe was confirmed prior to each measurement by a CCD camera equipped with $100\times$ magnifying objective. Probe fluence was 200 $\mu$J/cm$^2$ while the pump fluence varied in the region of $17 \div 170$ mJ/cm$^2$. The reflected probe pulse passed through linear \textit{s}-polarizer, in order to reduce scattered \textit{p}-polarized pump background, and was detected by Si detector. Another Si detector was used to monitor pulses intensity reflected from a glass wedge located before the sample. Signals of both detectors were averaged by a two-channel digital oscilloscope, the signal reflected from the sample was normalised to the monitored laser intensity signal.  The remaining experimental details  can be found elsewhere \cite{Kaplan}.  Using this technique we are able to resolve  the reflectivity  changes of $\Delta R/R_0\approx0.01$ (defined as $\Delta R \equiv R(\tau_d) - R_0$, where $R_0$ is the absolute reflectivity and $\tau_d$ is the delay time between the pump and probe).

\begin{figure}[tbh]
\centering\resizebox{\columnwidth}{!}{\rotatebox{0}{\includegraphics{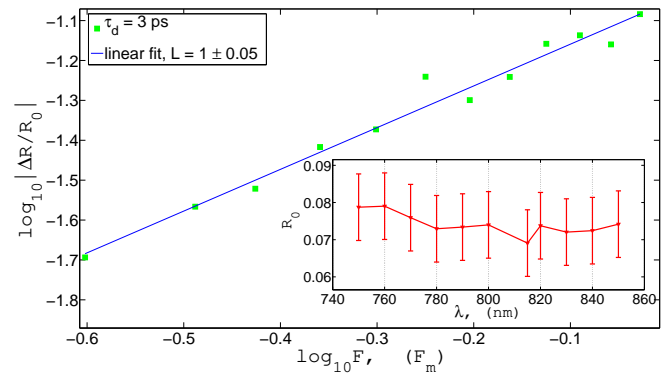}}}
\caption{\label{fig:3}(Color online) Normalized change of reflectivity versus fluence at delay time $\tau_{d}=3$~ps. Linear fit represents an order of the photoexcitation, $L=1\pm 0.05$. Note: supralinear scale. Inset: absolute reflection, $R_0$ in the vicinity of $\lambda=800$~nm. }
\end{figure}

A nano-imprint assistant dry-etching procedure has been employed to fabricate Si NP on Silicon-on-Insulator (SOI) structure by Lamontagne \textit{et al} \cite{sample}. The pillars are separated by $\sim125$ nm thick $SiO_{2}$ layer from the bulk Si substrate. The density of the NP in the nano-mask is $2.5 \times 10^{9}$ cm$^{-2}$. Fig.~\ref{fig:1} shows Scanning Electron Microscope (SEM) pictures of the sample with various resolutions. The highly-ordered structure of the NP is well resolved. The following average dimensions of the NP have been observed: diameter of $\sim70$ nm and height of $\sim200$ nm. The inset in Fig.~\ref{fig:3}  shows the absolute reflection, $R_0$, measurement  in the vicinity of $\lambda=800$ nm. According to this result,  $R_0= 0.075 \pm 0.01$ at  $\lambda = 800$ nm, roughly four times lower than $R_0$ of a crystalline Si. Similar value, of decreased $R_0$, was observed by G. R. Lin \textit{et al.} \cite{Lin} at an even broader spectrum on similar structures.

In the scope of the Drude model with neglected damping, the free carriers plasma refractive index is $n_p=n_0 \sqrt{1-\omega_p^2/\omega_0^2}$. Here $n_0$ and $n_p$ are the refractive indices of the bulk at the ground state and modified by the contribution from e-h plasma respectively. $\omega_0$ is the probing  frequency while the  plasma frequency denoted as $\omega_p = \sqrt{N_{e-h}e^2/\epsilon_0 m_{eff}}$.  $N_{e-h}$ represents e-h plasma density, $m_{eff}=(1/m_e+1/m_h)^{-1}$ is the reduced effective mass of the e-h pair, $\epsilon_0$ and $e$ carry their usual meanings. Neglecting non-linear response, the relation between the change in the reflectivity and in the refractive index ($\Delta n = n_p-n_0$) at normal incidence can be expressed as: $\Delta R/R_0 =4 \Delta n/ (n^2-1)$. Thus, $\Delta R \propto \Delta n \propto N_{e-h}$ allowing  extraction of $N_{e-h}$  from the  $\Delta R/R_0$  experimental data.
\begin{figure}[tbh]
\centering\resizebox{\columnwidth}{!}{\rotatebox{0}{\includegraphics{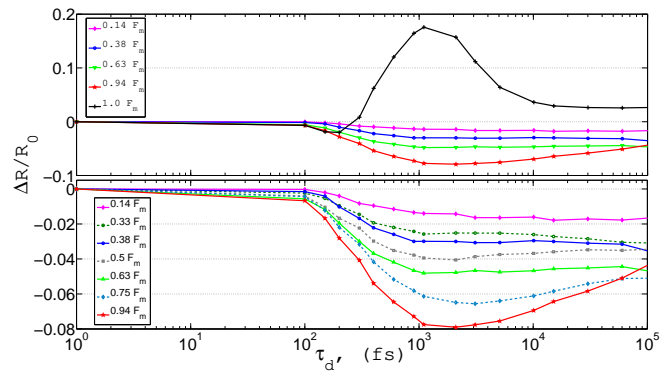}}}
\caption{\label{fig:2}(Color online) Transient reflectivity of silicon nanopillars as a function of the delay time  at various fluences values indicated in $F_m$ units. Upper panel: in the vicinity of melting fluence, $F_m$. Lower panel: below melting fluence. Some curves are shown on both panels for easier comparison. }
\end{figure}

The  time resolved  $\Delta R/R_0$ versus the delay time, $\tau_d$, recorded on the Si NP at  different fluences in the range of $0.14 \div 1 ~F_m$ (where $F_m = 170$ mJ/cm$^2$, is the threshold melting fluence  of Si surface \cite{Sokolowski}) is shown in Fig.~\ref{fig:2}. The upper panel shows that at the melting fluence, $F=1~F_m$, $\Delta R/R_0$ is predominantly positive. The initial reflectivity increase is attributed  to the metal-like optical response contributed mainly by high-density e-h plasma which determines  $n_p$ to be purely imaginary. After $\sim$1 ps $R(\tau_d)$ decreases  owing to carriers' collision-enhanced e-h recombination process. The rest of the curves, recorded at the fluences below the melting point, present negative $\Delta R/R_0$. Below the melting fluence a minimum of $\Delta R/R_0$  is reached after delay time of 1-2 ps. The largest change in the reflectivity, $\Delta R/R_0 = 8\%$, is observed  at $F = 0.94~F_m$.  For comparison, similar measurements near the threshold fluence on a crystalline Si wafer provided negative $\Delta R/R_0$  of $\sim1.5\%$ (not shown).
Fig. \ref{fig:3} shows a supralinear plot of $|\Delta R/R_0|$ versus $F$ at $\tau_d=3$ ps. From the plot, the coefficient of non-linearity, $L$, has been estimated  as $1 \pm 0.05$, demonstrating primary linear response of the pillars to the excitation.

\begin{figure}[tbh]
\centering\resizebox{\columnwidth}{!}{\rotatebox{0}{\includegraphics{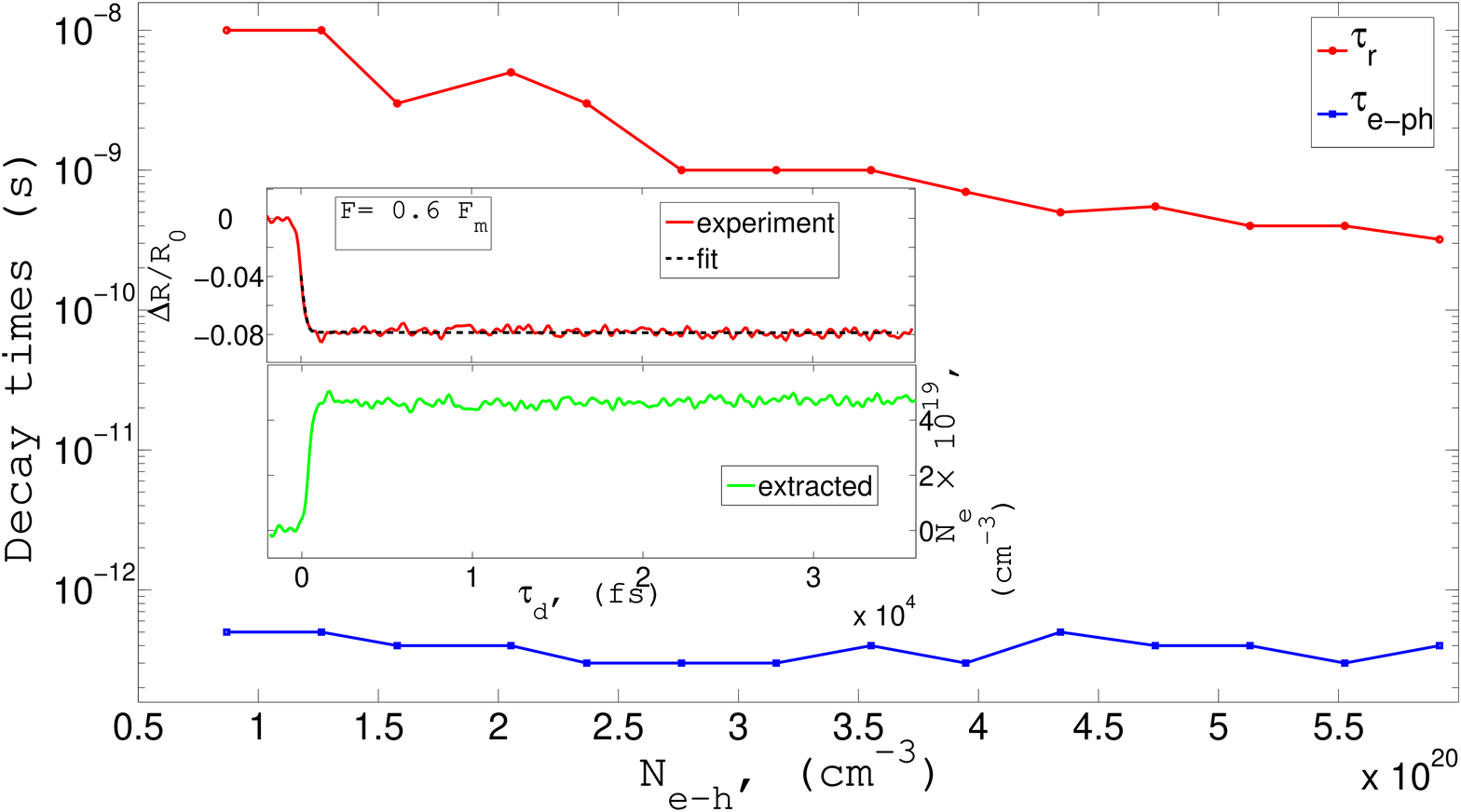}}}
\caption{\label{fig:4}(Color online) The extracted decay times vs plasma density, $N_{e-h}$. Blue curve:  the electron-phonon relaxation time, $\tau_{e-ph}$. Red curve: the electron-hole recombination time, $\tau_{r}$. Inset, upper part: experimental $\Delta R/R_0$ vs $\tau_d$ at $F=0.6 ~F_m$ and free carriers model fit. Inset, lower panel:  $N_{e-h} $ vs $\tau_d$ is extracted from $\Delta R/R_0$ in the scope of the free carriers absorption model.}
\end{figure}

In order to analyze the fast and the slow electron dynamics, we have adopted the two-exponential model presented by Sabbah \textit{et al.} \cite{Sabbah2002} (see Eq's 6-8), here we  note that for fitting we used $\Delta R(\tau_d)/R_0=\Delta R(0^+)/R_0 \times [\textrm{erf}(\tau_d/\tau_p)+1]/2 \times[1+(m_{eff}^{2000K}/m_{eff}^{300K}-1) \exp(-\tau_d/\tau_{e-ph})]^{-1} \times[1+C \exp(-\tau_d/\tau_r)]/(1+C)$, where $C$ is a fitting parameter. According to the model the fast component describes electron-phonon interaction and the slow represents e-h recombination. On the experimental curves, below the melting fluence, the fast component is responsible for the initial decrease in the reflectivity, $R(\tau_d)$, while the slow reflects its long recovery.  The upper inset of Fig.~\ref{fig:4} shows representative fitting of the model to the experimental  $\Delta R/R_0$ at $F= 0.6~F_m$. This experimental data was taken with higher resolution than that presented in Fig.~\ref {fig:2}. Similar analysis has been done for the decay curves measured at different fluences  and the decay constants have been extracted. Fig.~\ref{fig:4} presents these constants as a function of the initial $N_{e-h}$ which has been estimated assuming linear absorption with the coefficient $\alpha=1000$ cm$^{-1}$ and $R_0$ is taken from the inset of Fig.~\ref{fig:3}: $N_{e-h}=\alpha(1-R_0)F/E_{ph}$, where $E_{ph}$ is the photon excitation energy. 
The long decay time, $\tau_{r}$, describes the e-h recombination time. It  is found to depend strongly on the plasma density, decreasing  from 10 ns down to 300 ps, at the density range of $N_{e-h} = (0.8 \div 6)\times 10^{20}$  cm$^{-3}$. The lower inset of Fig.~\ref{fig:4} shows stability of the carriers population (after initial change) on the timescale of tens of picoseconds at a fluence of $F= 0.6~F_m$  ($N_{e-h} = 3.6 \times 10^{19}$ cm$^{-3}$). The short decay,  $\tau_{e-ph}$, is found to be independent of $N_{e-h}$ and estimated to be $\sim 350$ fs. 

In conclusion, we have studied carriers dynamics in the excited crystalline Si NP. It is clearly shown that under strong diffusion suppression and at near submelting regime, the recombination time of the e-h plasma is much longer than reported for a bulk material:  by UV excitation \cite{Tanaka} and diffusion dominated behavior \cite{Shank}. The thermalization time, however, remains unaffected. The initial drop of the reflectivity is up to 8\% which is  a few times higher than for the bulk silicon.

The authors appreciate the help of D C Sirica Ltd with the samples provided and thank E. Baskin for fruitful discussions during this work. We thank M. Y. Song for the SEM analysis. We acknowledge the financial support of  AWM
"Science City" network and UK EPSRC.

\end{document}